\documentstyle[aps,preprint] {revtex}
\begin{document}
\draft
\tightenlines

\title{Multiplicity of species in some replicative systems}   
\author{Adam Lipowski}
\address{ Department of Mathematics, Heriot-Watt University,\\ 
EH14 4AS Edinburgh, United Kingdom\\
{\rm and}\\
Department of Physics, A.~Mickiewicz University,\\
61-614 Pozna\'{n}, Poland}
\maketitle
\begin {abstract}
In an attempt to explain the uniqueness of the coding mechanism of living cells as contrasted
with the multi-species structure of ecosystems we examine two models of individuals with some
replicative properties.
In the first model the system generically remains in a multi-species state.
Even though for some of these species the replicative probability is very high, they are unable to
invade the system.
In the second model, in which the death rate depends on the type of the species, the system
relatively quickly reaches a single-species state and fluctuations might at most bring it to yet 
another single-species state.
\end {abstract}
\pacs{PACS 87.10.+e}
\newpage
\section{Introduction}
The problem of the origin of life and its early evolution is clearly one of the most
fundamental problems of modern science.
In spite of considerable efforts, even most basic questions in this multidisciplinary issue remain
unanswered.
The basic frame of most theories of the emergence of life was set many years ago by
Oparin~\cite{OPARIN}, who proposed that life emerged as a result of gradual evolution of
nonorganic matter.
Oparin's ideas were to some extent verified by experiments done by Miller and Urey~\cite{MILLER}, 
who showed that for a system of water and an atmosphere consisting of gases that were thought to be
common on the pre-biological Earth, electrical discharges resulted in the formation of some amino
acids and nucleotides.
It is believed that once created in sufficient concentrations, these molecules entered complicated
synthetic reactions, which produced more and more complex molecules.
Some of these complex molecules had catalytic and presumably to some extent even autocatalytic
properties~\cite{KAUFFMAN}.
The autocatalytic molecules (or rather systems of them), if of sufficient stability, were
clearly more likely to survive  in a competition for reactants.
The gradual evolution of such  autocatalytic systems, subjected to Darwinian selection, resulted in
mastering their surviving skills and eventually led to the emergence of life.
Even though the above-sketched scenario might seem plausible, many of its important details are
still unresolved~\cite{JOYCE}.

The replicatory mechanism of contemporary living cells is very sophisticated and basically
unchanged since the emergence of the first living cells, which presumably took place
about 3.5-4 billion years ago.
One of the characteristic features of this mechanism is that the tasks of coding and catalysis are
being assigned to different macromolecules, namely to nucleic acids and proteins, respectively.
Moreover, the code, i.e., the way amino acids are encoded by nucleotides, is universal
for all living cells.
It suggests an interesting possibility that all living cells are actually descendants of a single
pra-cell, which happened to develop the most effective surviving skills.

At the same time, however, it raises some questions.
One might expect that in search for the most effective cells Nature tried many variants of
different effectiveness.
Why did a certain code predominated all other variants?
Was this variant really of such an enormous effectiveness or maybe predomination was somehow a
generic feature of prebiotic dynamics?
One can notice, however, that in any contemporary ecosystem a large number of species coexist and
these species are clearly of different effectiveness.
Nevertheless the invasion of an ecosystem by a single species is an unobservable phenomenon.
Although very much different, both contemporary ecosystems and primeval soup might be regarded as
composed of certain individuals with some replicative properties.
Why thus did Nature select the single-species solution at the early stages of life and why
does it prefer multi-species solutions at later stages? 

Various aspects of the problem of the origin of life and biological evolution have been already
modeled~\cite{ROWE}.
Even though such models are, by necessity, highly simplified, they help us to understand the
essence of these complex phenomena.
For example, one can construct simple models of biological evolution which explain why the dynamics
of extinction of species has some scale-invariant properties~\cite{BAK}.

The problem of multiplicity of species in replicative systems has been also already addressed in
the literature.
A comprehensive review of bio-chemical aspects of this problem was written by Orgel~\cite{ORGEL} and
most recently by Szathm\'{a}ry~\cite{SZAT1}.
Certain simple models of replicative systems have been also recently examined~\cite{SZAT2,LIFSON}.
For example, in the model discussed by Szathm\'{a}ry and Maynard Smith an ensemble of
replicators is described in terms of differential equations.
In particular, the concentration of the $i$-th replicator is described by the following 
equation~\cite{SZAT1}:
\begin{equation}
dx_i/dt=k_ix_i^p, \label{szat}
\end{equation}
where $k_i$ is the growth rate constant of the $i$-th replicator and $p$ is the order of
replication.
It turns out that asymptotic ($t\rightarrow\infty$) concentrations (or rather their ratios) depend
on $p$.
For $p=1$ (Malthusian growth), the replicator with the largest $k_i$ becomes dominant and such a
case
is characterized as 'survival of the fittest'.
For $p<1$ (parabolic growth), the ratios of concentrations become finite, which is termed
as 'survival of everybody'.
In the case $p>1$ (hyperbolic growth), the dominant replicator is the one with the largest
product of the initial
concentration and the growth rate constant, which is termed as 'survival of the common'.
The interest in growth laws with $p\neq 1$ is partially motivated experimentally, since it was
shown
that certain oligonucleotides, which presumably played an important role in prebiotic dynamics, 
indeed follow the growth law with $p<1$~\cite{KIEDROWSKI,ZIELINSKI}.
Since such a growth implies 'survival of everybody', we are faced with a problem of the transition
to the Malthusian growth, which would explain 'survival of the fittest'.
However, recently it was shown by Lifson and Lifson that for more general growth laws
than Eq.~(\ref{szat}), the 'survival of the fittest' takes place even in the $p<1$
case~\cite{LIFSON}.

An important assumption underlying models leading to differential equations like
Eq.~(\ref{szat}) is that replicators are perfect, i.e., a replicator produces at a certain speed its
exact copy.
In our opinion, to model evolution at early stages we should rather consider a system of imperfect
replicators, which, for example, would produce their copies only with a certain probability and
otherwise they would mutate.

In the present Paper, we examine two simple models of such systems.
In our models, replicators, which might replicate or mutate, exist in infinitely many
varieties~\cite{COM1}.
As our main result, we show that behaviour of these models strongly depends on some details of
dynamics of these models.
Namely, only in one of these models the evolution proceeds in a single-species way, i.e., with the
majority of the system descending from the same ancestor.
In the second model, such single-species states are very unstable and the system evolves through 
multi-species paths.

In section II we define our models and present their basic properties.
In section III we examine in more details the behaviour of each model, emphasizing the difference
between single- and multi-species evolution.
Section IV contains our conclusions.
\section{Models and their basic properties}
\subsection{Model I}
Let us consider a system composed initially of $L$ individuals.
These individuals might be regarded as complex molecules at the prebiotic era immersed in the 
primeval soup and thus involved in a number of catalytic or autocatalytic reactions.
With each individual we assign randomly the replication probability $p_i \ (0<p_i<1$ for
$i=1,2\ldots,L)$ that the $i$-th individual will exactly replicate itself in the course of
reproduction.
The dynamics of this model, which in the following will be referred to as Model I, is specified
as follows:
\begin{enumerate}
\item {Choose an individual at random.
The chosen individual is denoted by $i$}.
\item {With the probability $L/N$ the $i$-th individual dies. 
The constant $N \gg 1$ might be regarded as a certain 'environmental capacity'.
Namely, provided that initially we have $L\leq N$, $L$ will never exceed $N$.}
\item {With the probability $1-L/N$ the $i$-th individual survives and produces
a new individual $j$.
The probability $p_j$ assigned to the $j$-th individual is equal to $p_i$ (parent's value) with
the probability $p_i$ and is equal to a random number from the
interval (0,1) with the probability $1-p_i$.
This rule means that if a copying error happens, it dramatically changes the properties of
the new individual.}
\end{enumerate}
These rules imply that in the steady state the death rate $(L/N)$ equals to the
reproduction rate $(1-L/N)$ and thus on average $L=N/2$.
As far as the number of individuals is concerned, Model I is equivalent to a certain random walk
problem.
Indeed, from the rules stated above one can infer the following equation for the probability
$P(L,t)$ that
in our system there are $L$ individuals at time $t$:
\begin{equation}
P(L,t+\Delta)=\frac{L}{N}P(L+1,t)+(1-\frac{L}{N})P(L-1,t).
\label{1}
\end{equation}
This equation describes changes in our system after updating a single molecule.
To conform to the Monte Carlo simulations presented later, we assume that such a single update
takes $\frac{1}{L}$ of time (i.e., a unit of time corresponds to a single, on average, update of
each individual).
Thus, in Eq.(\ref{1}) we have $\Delta=\frac{1}{L}$.
Introducing the variable $M=L-\frac{N}{2}$, we can rewrite Eq.(\ref{1}) as
\begin{equation}
P(\frac{N}{2}+M,t+\Delta)=(\frac{1}{2}-\frac{M}{N})P(\frac{N}{2}+M-1,t)+(\frac{1}{2}+ 
\frac{M}{N})P(\frac{N}{2}+M+1,t),
\label{2}
\end{equation}
which is clearly the equation of a random walk with attraction toward $M=0$, i.e., $L=\frac{N}{2}$.
It has been already shown that the so-called 'dog-flea' model is also equivalent to a similar
random
walk problem with attraction~\cite{KAC} and that fluctuations in this model around equilibrium
($M=0$) become negligible in the limit, which in our case corresponds to $N\rightarrow \infty$.
Thus, we expect that in Model I fluctuations of the number of individuals around $L=\frac{N}{2}$
are also small for large $N$.

To examine replicative properties of our model, we resort first to Monte Carlo simulations.
Since the implementation of the above rules on the computer is rather straightforward, we present
only the results of these simulations.
In Fig.\ref{f1} we present the time evolution of the average replication probability
$p(t)=\frac{1}{L}\sum_{i=1}^{L} p_i$.
Simulations were done for $N=10^7$ and initially the probabilities $p_i \ \ (i=1,\ldots,N)$ were
chosen at random.
In all simulations reported in the present paper the initial number of individuals $L$ is equal to
$N$.
One can clearly see that the average replication probability increases in time
but the increase is very slow and it is not obvious what value is reached in the steady state
(i.e., for $t \rightarrow \infty$).

However, below we present some analytical calculations which show that if the limit
$N\rightarrow\infty$ is taken first, then for $t\rightarrow\infty$  the average replication
probability $p(t)$ converges to unity.
Our strategy is to write the evolution equation for the higher order moments of replication
probability  and then to solve the resulting infinite set of equations in the steady state.
First let us assume that the evolution in our model lasted long enough to equilibrate it with
respect to the number of molecules $L$.
Thus, we approximate the death and reproduction probabilities as
$\frac{L}{N}=1-\frac{L}{N}=\frac{1}{2}$.  
Introducing the notation $p^l(t)=\frac{1}{L} \sum_{i=1}^L p_i^l$, where $l=1,2\ldots,$ we can write
the following evolution equation for the first moment of $p_i$ (i.e., for $p(t)$) 
\begin{equation}
\frac{N}{2}p^1(t+\Delta)- 
\frac{N}{2}p^1(t)=-\frac{1}{2}p^1(t)+\frac{1}{2}[p^2(t)+\frac{1}{2}(1-p^1(t))],
\label{3}
\end{equation}
were we put $\frac{N}{2}$ as an average number of individuals.
The first and second terms on the right-hand side of Eq.(\ref{3}) describe the changes due to a
single
update caused, respectively, by the death and reproduction processes.
The term $\frac{1}{2}(1-p^1(t))$ corresponds to production of an individual with a randomly
assigned
replication probability, which thus on average takes the value $\frac{1}{2}$. 
We can write similar equations for arbitrary $l$:
\begin{equation}
\frac{N}{2}p^l(t+\Delta)-\frac{N}{2}p^l(t)=
-\frac{1}{2}p^l(t)+\frac{1}{2}[p^{l+1}(t)+\frac{1}{l+1}(1-p^1(t))], \ \ {\rm for} \
l=1,2\ldots \ .
\label{4}
\end{equation}
In Eq.(\ref{4}) we used the fact that the $l$-th moment of a random variable, which is uniformly
distributed on (0,1), is equal to $\int_0^1 s^l ds =\frac{1}{l+1}$.
In the steady state, the left-hand side of Eq.(\ref{4}) is zero and thus we obtain
\begin{equation}
p^l=p^{l+1}+\frac{1}{l+1}(1-p^1) \ \ {\rm for}\ \  l=1,2\ldots \ .
\label{5}
\end{equation}
This infinite set of equations can be solved.
Namely, when we add the Eqs.(\ref{5}) for $l=1,2,\ldots,$ all higher-order moments cancel out and we
obtain
\begin{equation}
p^1=(1-p^1)\sum_{i=2}^{\infty} \frac{1}{i},
\label{6}
\end{equation}
and thus $p^1=\sum_{i=2}^{\infty} \frac{1}{i}/\sum_{i=1}^{\infty} \frac{1}{i}=1$, since both series
diverge and differ only by unity.
Similarly, all other moments $p^l$ in the steady state are equal to one.
Thus we expect that in the limit $t\rightarrow\infty$ the replicative probability $p(t)$
in Fig.\ref{f1} increases to unity even though the convergence seems to be very slow.
\subsection{Model II}
Before discussing other properties of Model I, let us consider the model where the probability of
death of a certain individual depends not only on the total number of individuals (as in Model I)
but also on the individual itself.
Such a modification is motivated by the fact that in the primeval soup the survival of a molecule
was determined not only by the access to substrates (and then the total number of molecules is
likely to determine the death rate) but also by the stability of a given molecule against, e.g.,
radiation, which clearly depends on the type of this molecule.

Thus, let us consider the model which in the following will be referred to as Model II.
To each individual $i$, in addition to the replication probability $p_i$ we assign randomly certain
individual survival probability $r_i$ ($0<r_i<1$).
The dynamics of this model is specified
as follows:
\begin{enumerate}
\item {Choose an individual at random.
The chosen individual is denoted by $i$.}
\item {With the probability $L/N$ the $i$-th individual dies due to the lack of reproductive
substrates.}
\item {Provided that the individual survived the previous step:}
\begin{itemize}
\item{it dies with the probability $1-r_i$,}
\item{it survives with the probability $r_i$ and reproduces according to the rule analogous to
that of Model I.
Namely, the new individual with the probability $p_i$ has the same replication probability and
death probability as its parent (i.e., $p_i$ and $r_i$, respectively) and with the probability
$1-p_i$ these probabilities are chosen randomly anew.} 
\end{itemize}
\end{enumerate}

For $r_i=1 \ \ (i=1,\ldots, L)$ Model II becomes equivalent to Model I.
Monte Carlo simulations for Model II are also straightforward and we present only the results.
In Fig.\ref{f1} we present the average replication probability $p(t)$ defined in the same way as
for
Model I.
One can see that the convergence to unity is in this model much faster than in Model I.
\section{Single- versus multi-species evolution}
But there are more important differences between these models than the rate of convergence.
Certain indication of different behaviour is seen in Fig.\ref{f2}, where we plot $1-p(t)$ as a
function
of time in a logarithmic scale.
One can see that the late-time evolution of Model II proceeds in steps between which the system
basically remains at the same level of $p(t)$ and  none indication of such a stepwise behaviour is
seen for Model I.
This stepwise behaviour suggests that Model II remains mostly in a single-species state with
majority of individuals belonging to the same species.
Individuals $i$ and $j$ belong to the same species if $p_i=p_j$ and $r_i=r_j$; for Model I the
second condition is always satisfied.

To confirm  such a scenario we present in Fig.\ref{snap} snapshot configurations for Models I and
II after 5000 Monte Carlo steps and $N=10^5$.
Indeed, after this simulation time Model II was brought to a single-species state with
only few individuals of $p_i\neq 0.9996\ldots$ (they are not shown in Fig.\ref{snap} since they all
have $p_i<0.99$).
On the other hand, Model I still remains in the multi-species state.

We would like to point out that, of course, for finite $N$ there exists a finite probability that
Model I can be brought into a single-species state (and for small $N$ one can indeed see such a
behaviour in simulations) but for large $N$ ($\sim 10^5$) this would require an extremely long
simulation time.
What is, however, more important is that for Model I single-species states are relatively unstable
and this model mostly remains in multi-species state. 
Such a behaviour is clearly seen in Fig.\ref{domin}, where we show the time evolution of the
percentage $s$ of individuals belonging to a dominant species in the system.
As an initial configuration we have chosen a single-species state with prescribed values of
replication and death probabilities, namely we set $p_i=r_i=p_0$, where $p_0=0.9$ or $0.99$ for
$i=1,2,\ldots,N$ (for model I, $r_i=1$ independently on $p_0$).
One can see that Model I indeed quickly abandons the single-species state.
We have checked that also for a larger $p_0$ the behaviour of Model I is basically the same and the
model quickly evolves toward the multi-species state.
Evolution of Model II is different.
When prepared in a state of large $p_0$ (e.g., 0.99), the model remains in this state for very long
time.
Even when $p_0$ is smaller (e.g., 0.9), this model, after some short transient, ends up in a
single-species state.
Further evolution of Model II consists of small fluctuations within such a state, which sometimes
might be strong enough to bring the system, again via some short transient, to another
single-species state and usually with larger replication and survival probabilities.

Such a scenario is also confirmed in Figs.\ref{fdod1}-\ref{fdod3}.
We simulated the system of the size $N=2\cdot 10^4$ and with random initial probabilities $p_i$ and
$r_i$.
In Fig.\ref{fdod1}, which shows the occupation of dominant species $s$, one can see that Model I has
rather irregular behaviour.
Sometimes a dominant species occupies a great majority of the system ($s\sim 1$) but sometimes it
is only a small fraction of the system.
The behaviour of Model II is different (Fig.\ref{fdod2}).
The dominant species almost always occupies most of the system.
Only during very rare and short-lived fluctuations, $s$ becomes substantially smaller than unity.
In Fig.\ref{fdod3} we show the replication probability of the dominant species for the
Monte Carlo runs shown in Fig.\ref{fdod1} and Fig.\ref{fdod2}.
One can see that in Model II, contrary to Model I, the dominant species are very longed-lived.
A comparison of Fig.\ref{fdod3} with Fig.\ref{fdod1} shows that even when the dominant species is
unchanged, the percentage $s$ occupied by it might fluctuate wildly.
\section{Conclusions}
We have examined two very simple models of systems of replicative individuals.
Although both models seem to evolve toward the state of perfect replicability, their evolution is
markedly different.
For Model II evolution proceeds through a sequence of consecutive transitions, between
which the system remains basically in a single-species state (i.e, with almost all individuals
being identical).
In our opinion, this model might describe prebiotic evolution until the invention of the universal
code.
According to this model, when the relatively stable and almost error-free replicative mechanism was
found, it quickly invaded the whole system.
On the other hand, Model I during its evolution remains mostly in a multi-species state.
Such a multi-species structure resembles modern ecosystems, where a large number of species
coexist and are constantly struggling for survival.

Finally, let us notice that although the evolution of a system as a whole seems to proceed slower
in Model I than in Model II (Fig.\ref{f1}), there are some species in Model I with replicative
probabilities very close to unity (Fig.\ref{snap}, Fig.\ref{fdod3}).
Thus, multi-species dynamics in Model I enhances nucleation of species of very high effectiveness.
However, even they are unlikely to 'reign' for a long time.

\begin{figure}
\caption{The average replication probability $p(t)$ as a function of time $t$ for Model I
(solid line) and Model II (dotted line).
Simulations were performed for $N=10^7$ and initially the probabilities $p_i, \ \ (i=1,\ldots,N)$
were chosen at random.} 
\label{f1}
\end{figure}
\begin{figure}
\caption{The logarithmic plot of $1-p(t)$ as a function of $t$ for $N=10^5$ and initial
conditions as in Fig.\ref{f1}.}
\label{f2}
\end{figure}
\begin{figure}
\caption{The replication probability $p_i$ as a function of $i$ for Model I (small dots)
and Model II (diamonds) after 5000 Monte Carlo steps and for $N=10^5$.
Almost all individuals for Model II have the same value of $p_i=0.9996\ldots$ and the corresponding
diamonds constitute a thick line in the upper part of the figure.}
\label{snap}
\end{figure}
\begin{figure}
\caption{ Occupation of a dominant species $s$ as a function of time for
Model I (bottom lines) and Model II (top lines).
In the initial configuration all individuals are identical with the replication and
death probabilities equal to 0.9 or 0.99.
Simulations were done for $N=2\cdot 10^4$.}
\label{domin}
\end{figure}

\begin{figure}
\caption{The occupation of the dominant species $s$ as a function of time for Model I.
Simulations, were done for $N=2\cdot 10^4$ and random initial probabilities $p_i \ \
(i=1,2,\ldots,N)$.}
\label{fdod1}
\end{figure}

\begin{figure}
\caption {The occupation of the dominant species $s$ as a function of time for Model II.
Simulations were done for $N=2\cdot 10^4$ and random initial probabilities $p_i$ and $r_i \ \
(i=1,2,\ldots,N)$.
Three large fluctuations seen for $t<50000$ resulted in changing the dominant species
(see Fig.\ref{fdod3}).}
\label{fdod2}
\end{figure}

\begin{figure}
\caption{ The replication probability $p_i$ of the dominant species for Model I (dots) and Model
II (densely plotted $\diamond$ which constitute basically a thick line at $p_i \sim 0.997$).}
\label{fdod3}
\end{figure}


\begin {thebibliography} {00}

\bibitem{OPARIN} A.~I.~Oparin, {\it The origin of life on the earth}, (New York: Academic Press,
1957).

\bibitem{MILLER} S.~L.~Miller, Science {\bf 117}, 528 (1953).
S.~L.~Miller and H.~Urey, Science {\bf 130}, 245 (1959).

\bibitem{KAUFFMAN} It was suggested that autocatalytic properties are a generic feature of
a sufficiently complex mixture of proteins: S.~Kauffman, J.~Theoret.~Biology {\bf 119}, 1 (1986).

\bibitem{JOYCE} G.~F.~Joyce, The New Biologist {\bf 3}, 399 (1991).

\bibitem{ROWE} G.~W.~Rowe, {\it Theoretical Models in Biology}, (Clarendon Press-Oxford, 1994).

\bibitem{BAK} P.~Bak and K.~Sneppen, Phys.~Rev.~Lett.~{\bf 71}, 4083 (1993).
R.~V.~Sole and S.~C.~Manrubia, Phys.~Rev.~{\bf E 54}, R42 (1996).
L.~A.~N.~Amaral and M.~Meyer, Phys.~Rev.~Lett.~{\bf 82}, 652 (1999).

\bibitem{ORGEL} L.~E.~Orgel, Nature {\bf 358}, 203 (1992).

\bibitem{SZAT1} E.~Szathm\'{a}ry, Trends in Genetics, {\bf 15}, 223 (1999).

\bibitem{SZAT2} E.~Szathm\'{a}ry and J.~Maynard Smith, J.~Theor.~Biol.~{\bf 187}, 555 (1997).

\bibitem{LIFSON} S.~Lifson and H.~Lifson, J.~Theor.~Biol.~{\bf 199}, 425 (1999).

\bibitem{KIEDROWSKI} G.~von Kiedrowski, Angew.~Chem.~Int.~Ed.~Engl.~{\bf 25}, 932 (1986).

\bibitem{ZIELINSKI} W.~S.~Zielinski and L.~E.~Orgel, Nature {\bf 327}, 346 (1987).

\bibitem{COM1} It was already emphasized that biologically relevant replicators should potentially
exist in infinitely many forms~\cite{SZAT2}.

\bibitem{KAC} M.~Kac and J.~Logan, in {\it Fluctuation Phenomena}, eds. E.~W.~Montroll and
J.~L.~Lebowitz (North Holland Physics Publishing, 1987).

\end{thebibliography}
\end {document}
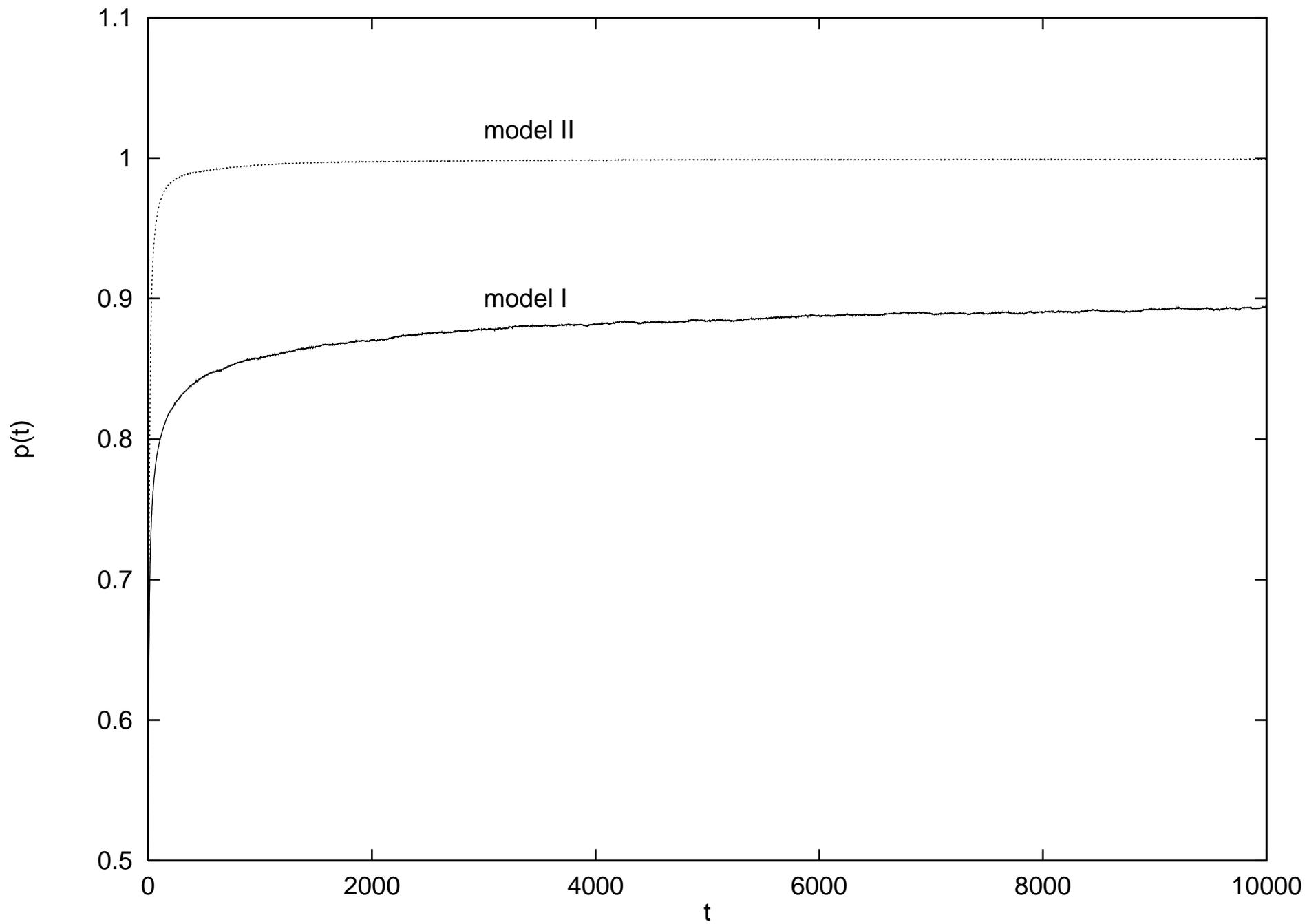
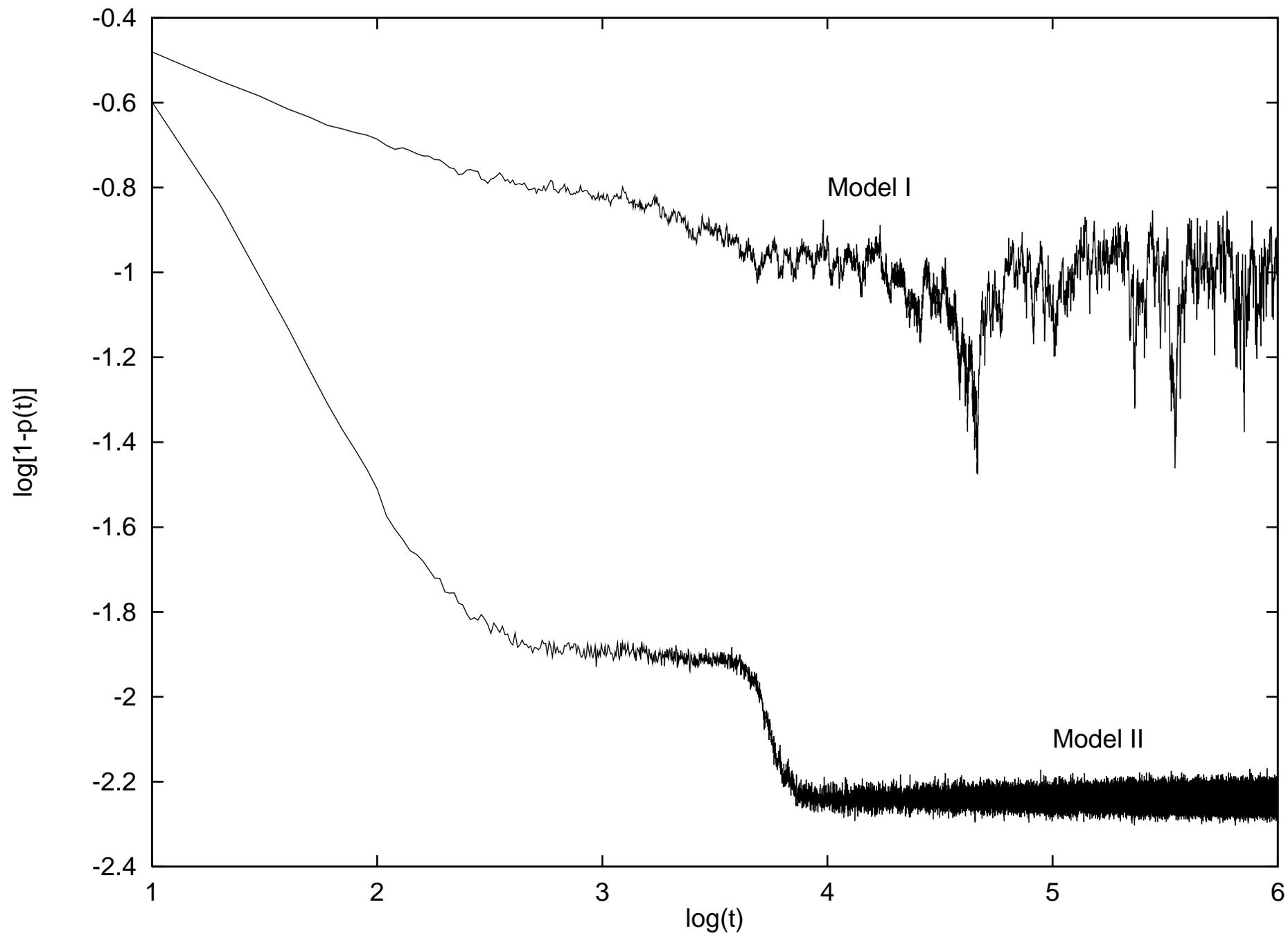
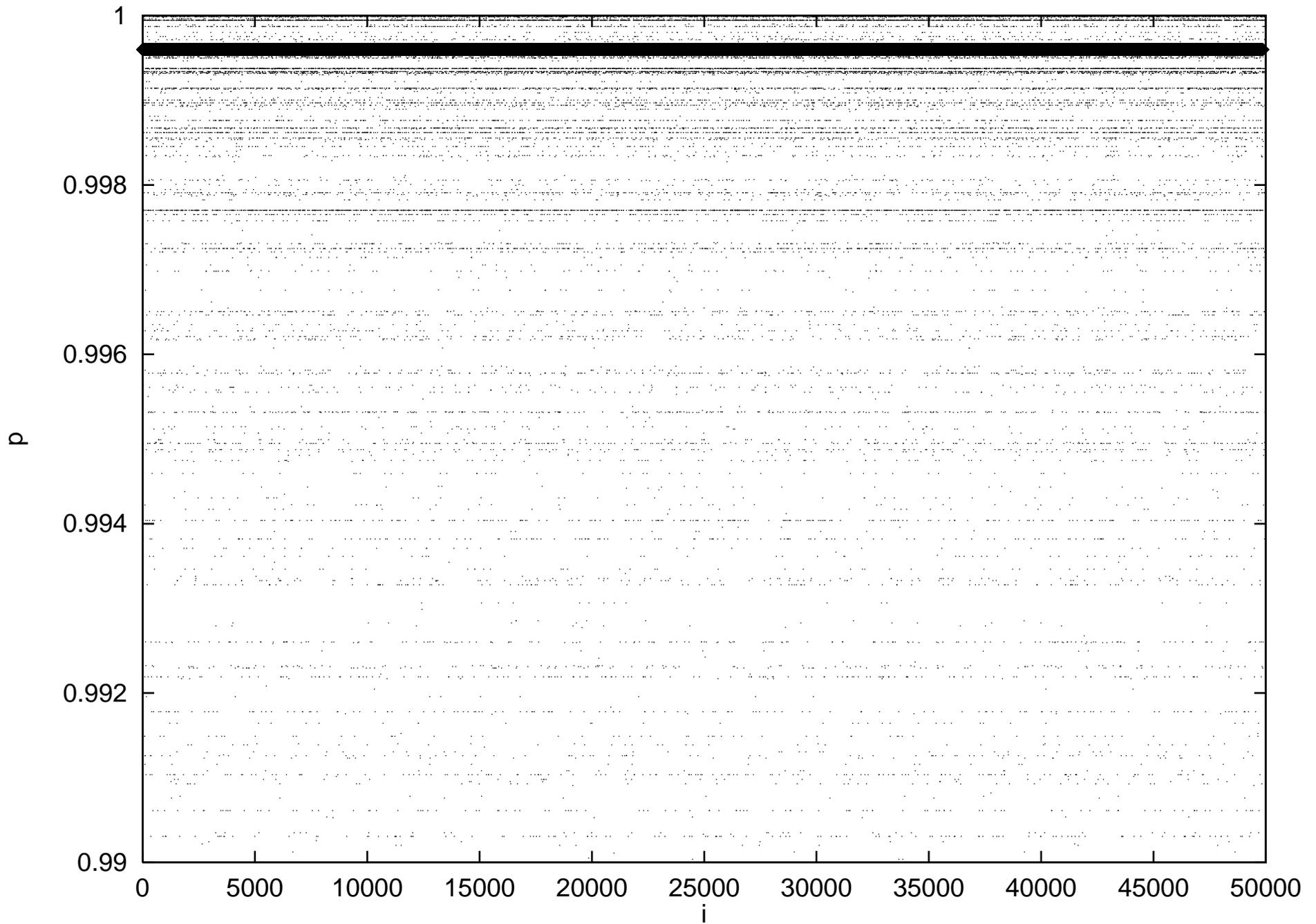
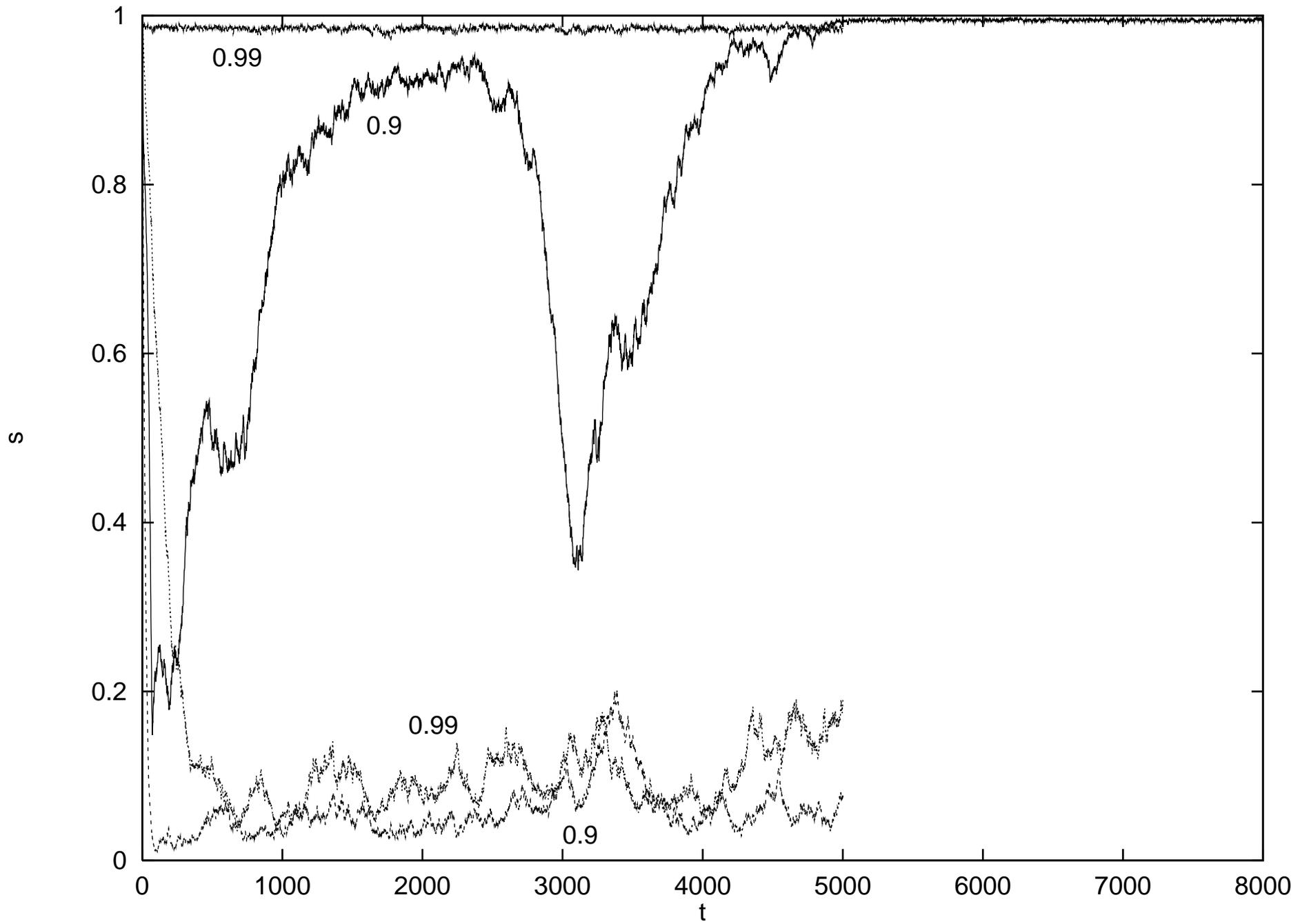
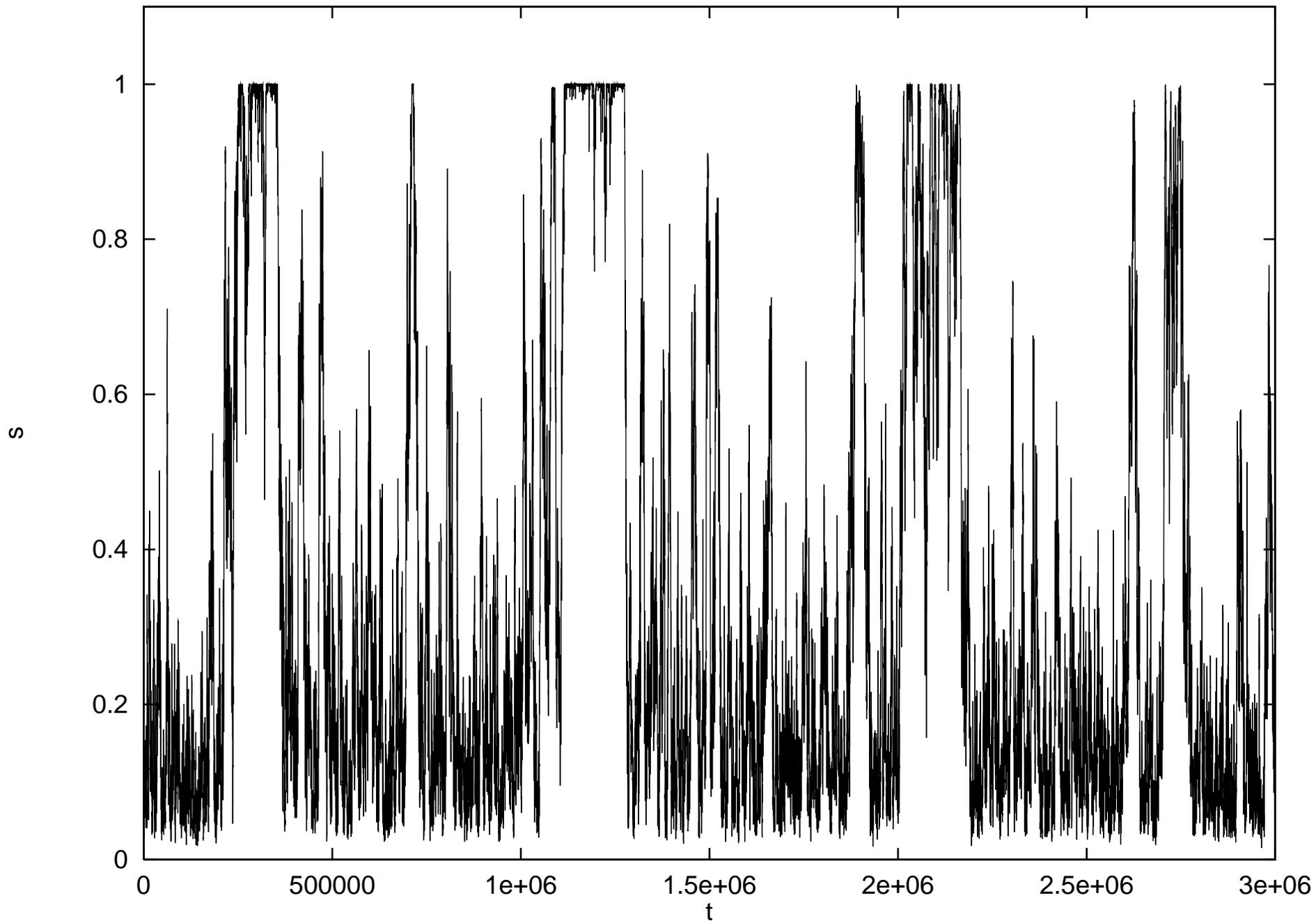
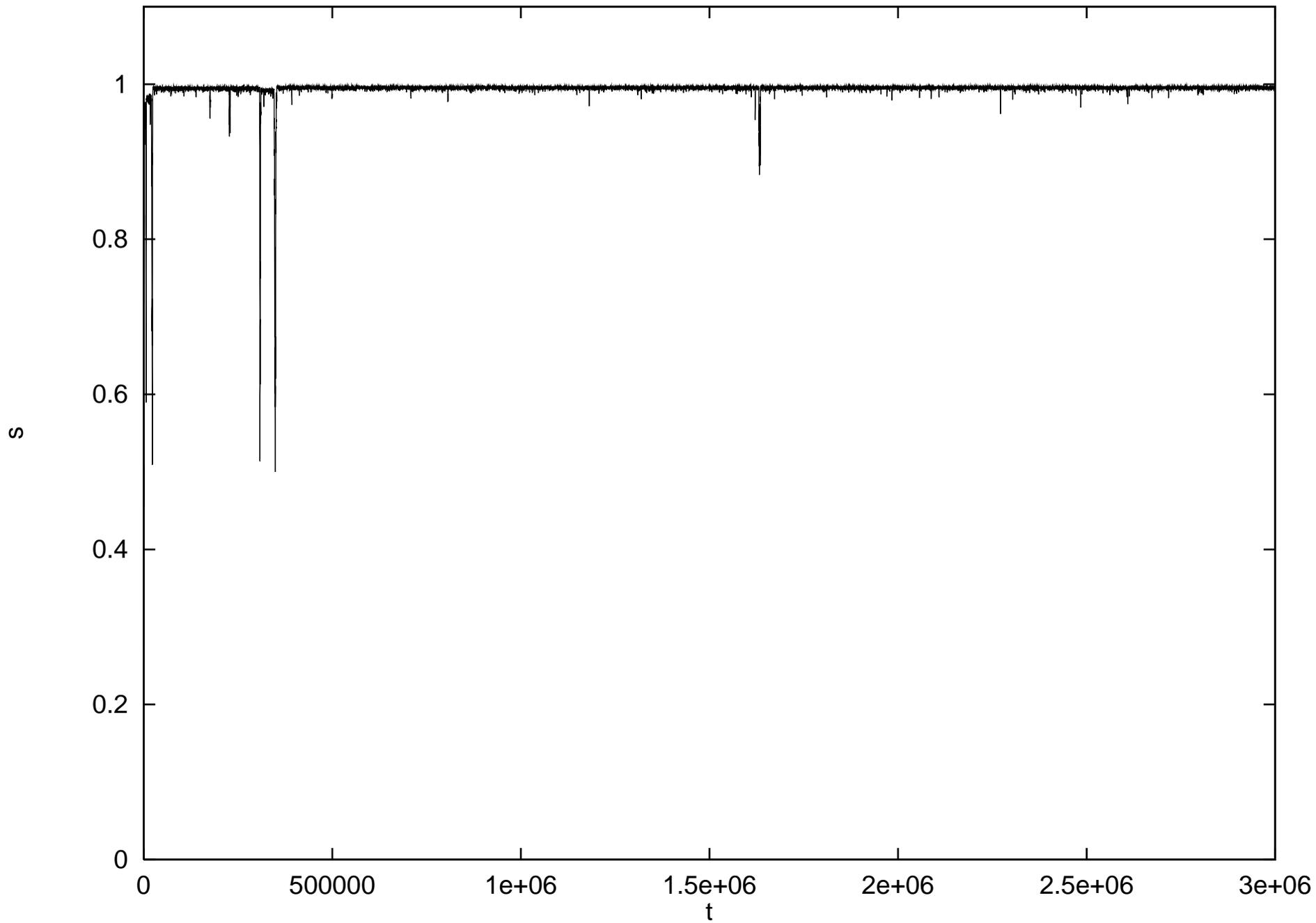
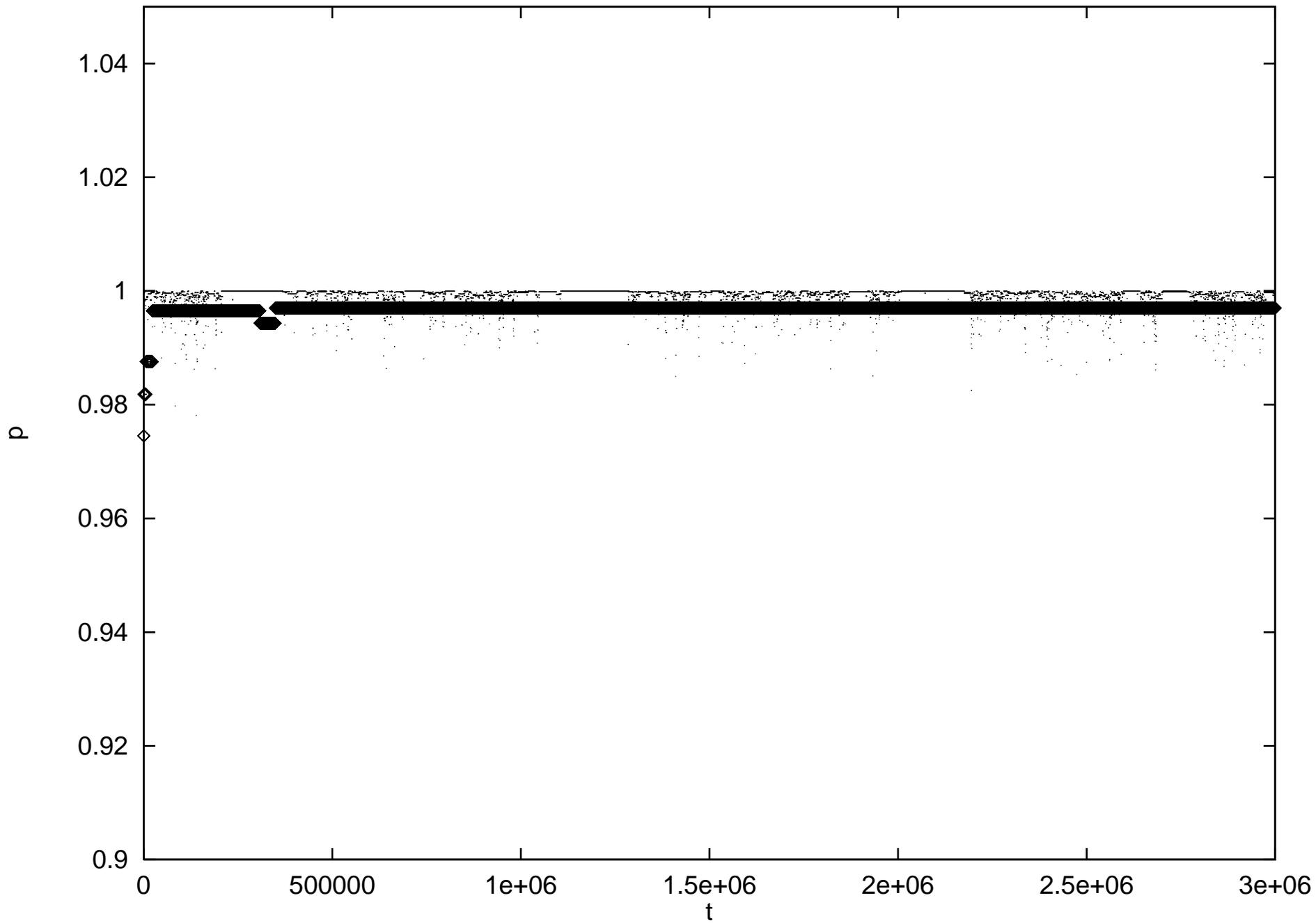
